\documentclass[aps,twocolumn,showpacs,superscriptaddress]{revtex4-1}
\usepackage{amsmath,amssymb}
\usepackage{graphics,graphicx}
\usepackage{dcolumn,bm}
\usepackage{psfrag}
\newcommand{\cu}
{\affiliation{Department of Physics, University of Calcutta, 
92 Acharya Prafulla Chandra Road, Kolkata 700009, India.}}

\begin{document}
 
\title{Effect of randomness in  logistic maps
}
\author{Abdul Khaleque}%
\cu
\author{Parongama Sen}%
\cu


\begin{abstract}
We study a random logistic map $x_{t+1} = a_{t} x_{t}[1-x_{t}]$ where 
$a_t$ are bounded 
($q_1 \leq a_t \leq q_2$), 
random variables independently drawn from a distribution. 
 $x_t$ does not show any regular behaviour in time. We find that $x_t$ shows fully ergodic behaviour
when the maximum allowed value of $a_t$ is $4$. 
However $\langle x_{t \to \infty}\rangle$, averaged over different realisations  
reaches a fixed point. 
For $1\leq a_t \leq 4$ the system shows nonchaotic behaviour and the Lyapunov exponent is strongly
dependent on the asymmetry of the distribution from which $a_t$ is drawn.
Chaotic behaviour is seen to occur  beyond a threshold
value of $q_1$ ($q_2$)
when $q_2$ ($q_1$) is varied. The most striking result is that the random map is chaotic even when $q_2$ is less than the threshold
value $3.5699......$ at which chaos occurs in the non random map. We also employ a different method in which a 
different set of random variables are used for the evolution of two initially identical $x$ values, 
here the chaotic regime exists for all $q_1 \neq q_2 $ values.

\end{abstract}


\maketitle
\section{Introduction}
Many natural phenomena show chaotic behaviour and a possible route to chaos is provided by nonlinear dynamical models.
 In the study of nonlinear dynamics, logistic map is a well known area of research \cite{strogat}.
Study of logistic map is relevant  in population dynamics \cite{may}, image encryption \cite{baptista,wong,pareeka}, electronic circuit, 
pseudo-random number generators \cite{phatak,vinod} etc. Mathematically, the logistic map is given by
\begin{equation}
 x_{t+1}=a x_t(1-x_t).
\label{eq-logistic}
\end{equation}
The logistic map is perhaps the most simple and illustrative example of iterative maps of the form $x_{t+1}=f(x_t)$
which shows very complex dynamical behaviour. In this map, for $1\leq a \leq 3$ a nonzero steady value is reached. For larger $a<a_c=3.5699......$
 multiple fixed point values with
 periodicity $2,4,6,.....$ etc. occur, while above $a_c$, chaos occurs which is exhibited by
a positive Lyapunov exponent \cite{strogat}.

In reality, dynamical phenomena e.g. population dynamics, usually involve an amount of stochasticity. In order to incorporate this feature, we analyse the behaviour of the logistic map where $a$ does not have a fixed value but is a random variable $a_t$, 
drawn independently from a distribution. In some earlier works \cite{cham,bhatt,stein}, such randomisation 
of $a$ has been considered where the main issue was to check whether ${x_t \to 0}$ as ${t \to \infty}$ such that the distribution
 of $x$ at large times is a delta function at $x=0$. It was concluded that under certain condition this is true.

We  use different distributions for choosing the control parameter $a_t$.   We choose $a_t$  to be bounded, i.e., 
 $q_1\leq a_t \leq q_2$ with $q_1 \ge 1$ to avoid the fixed point $x=0$ as obtained in the earlier cases \cite{bhatt,cham,stein}.

We study  the behaviour of 
the quantity $\Delta_t$ defined as 
\begin{equation}
 {\Delta }_t=\langle |x_t- x^\prime_t|\rangle,
\label{eq-delta-t}
\end{equation}
where $x_t$ and $x^\prime_t$ denote the two different evolutions. 
$\Delta_t$ is the average value obtained from different configuration as denoted
by the angular brackets on the right hand side of (\ref{eq-delta-t}). 
The evolution has been implemented in two different ways:
in the traditional method (TM), one starts with  two initially close values $x_0$ and $x^\prime_0$ to calculate $\Delta_t$ as these 
two initial $x$ values are evolved using the same set of $a_t$.
In the other method, different sets of $a_t$ are used for initially 
identical values of $x$. The latter is called the ``Nature vs. nurture'' method following 
\cite{machta}.  

As usual, if ${\Delta }_t$ grows (or saturates at a finite value  as it cannot increase indefinitely)
we conclude that the chaotic regime is reached. Varying $q_1$ and $q_2$ we identify such regions in both the methods.
Apart from identifying the chaotic region, we are also interested in comparing the 
TM and NVN methods which have led to different results in interacting dynamical systems \cite{machta,khaleque}.

One of the main objectives is to study whether universal behaviour is observed when we vary the distribution.
For this purpose, we have used a uniform distribution and  a  triangular distribution which has a peak at a value 
$a_p$ ($q_1 \leq a_p \leq q_2$), from which the $a$
values are chosen. For the  triangular distribution $a_p=(q_1+q_2)/2$ signifies a symmetric triangular distribution. 
Deviation of $a_p$ from this value quantifies the asymmetry.

In the next two sections we discuss the results and in section IV summary and discussions are presented. 
\section{Ergodicity and convergence}
We first report some general results which are in stark contrast to the non-random map. We use either a uniform
distribution or a symmetric triangular  distribution for these studies.

Let the initial value of $x$ be chosen randomly from [$0$:$1$].
In the non-random logistic map, it is known that $x$ has finite number of attractors in  the nonchaotic regime.
In the random case however $x$ has an ergodic type behaviour in the sense that 
it does not reach a fixed attractor and can assume values between $x_{min}$ and $x_{max}$; $\Delta_x=(x_{max}-x_{min})$
is nonzero for $q_1\neq q_2$. When $q_1$ is fixed at $1$,  $\Delta_x$ shows an increasing behaviour
with $q_2$. However when $q_2=4$,  $\Delta_x\simeq 1$ for any value of $q_1 \geq 1$ showing that the system becomes fully ergodic (fig.~{\ref{rawdyn_ergo}}).
The fate of two independent evolutions depends on $q_1$ and $q_2$; however we note
that the average value  $\langle x_t \rangle$  of all such evolution shows convergence and is non ergodic.

\begin{figure}[!h]
\resizebox{90mm}{!}{\includegraphics {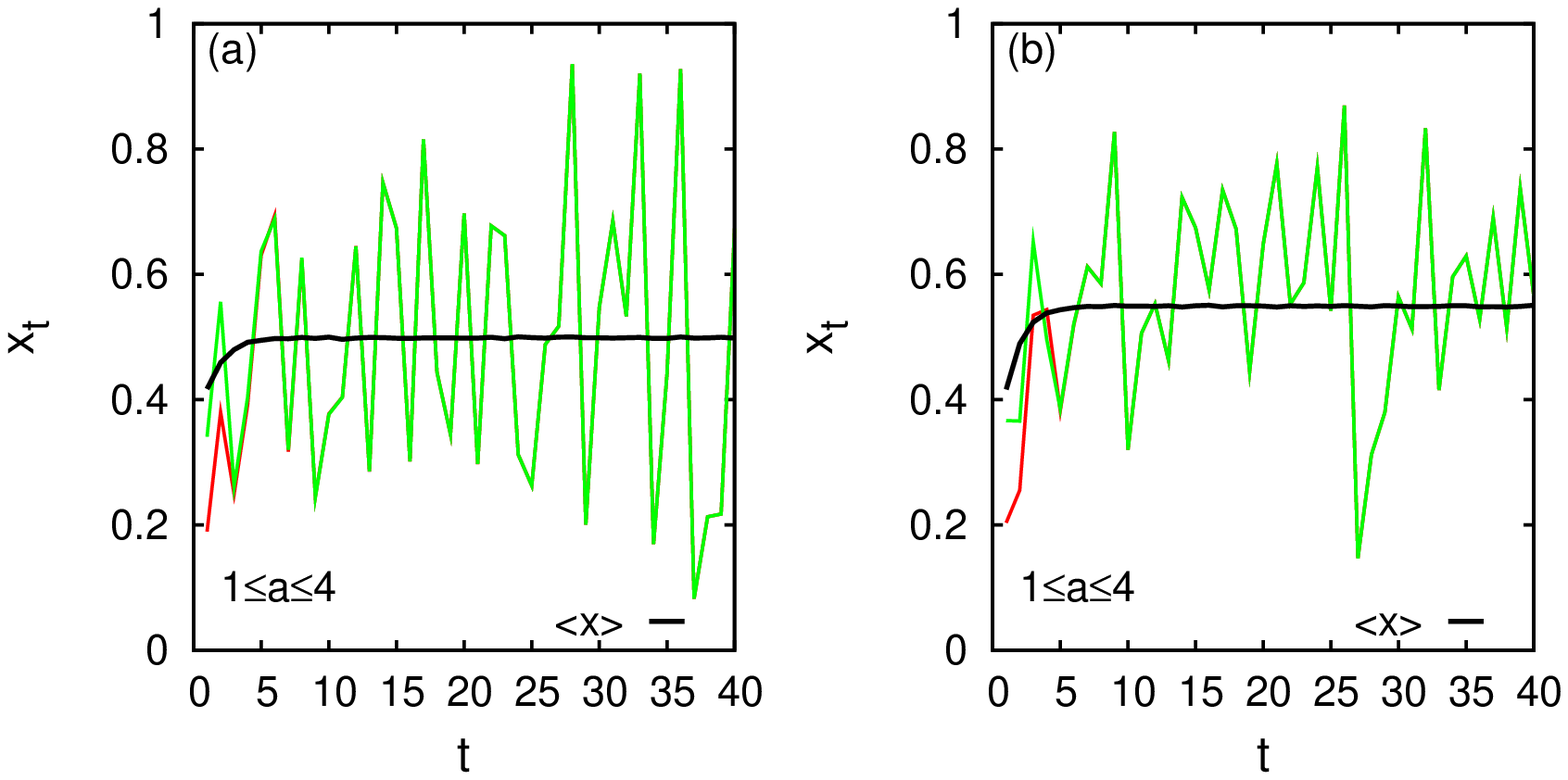}}

\caption{(Color online) TM results: Two different evolutions of $x(t)$  in the traditional method. Left panel: uniform distribution; Right panel: symmetric triangular distribution.
The steady state  values  averaged over many configurations also shown (see also figs.~{\ref{ndyn_in_tm}}a, {\ref{ndyn_in_tm}}b).
}
\label{rawdyn_tm}
\end{figure}

It is known that for a fixed value of $a$, the non-random logistic map has a fixed point at $x = \frac{a-1}{a}$ which is stable
below $a=3$. For the 
random case, one can  
approximate  theoretically,
\begin{equation}
 \langle x \rangle = \frac {\bar a -1}{\bar a},
\label{non_random}
\end{equation}
  where ${\bar a}$ denotes the average value. 
If we allow the distribution of $a$ to vary between $q_1$ and $q_2$,
we find that the value of $\langle x_{t \to \infty}\rangle$ obtained numerically differs from $\langle x \rangle$ obtained using eq.~{\ref{non_random} (Table \ref{TMT}).
We have shown that while the non ergodicity behaviour is present for both the uniform and symmetric triangular distribution, 
the deviation of $\langle x \rangle$ is larger for the uniform distribution compared to that in the symmetric triangular
 distribution from $\langle x_{t \to \infty}\rangle$.
 Also, ($\langle x \rangle-\langle x_{t \to \infty}\rangle)/\langle x \rangle$  
increases with $q_2-q_1$ in
the uniform distribution. However, this increase appears to be weaker in case of the symmetric triangular distribution.
\begin{figure}[!h]
\resizebox{90mm}{!}{\includegraphics {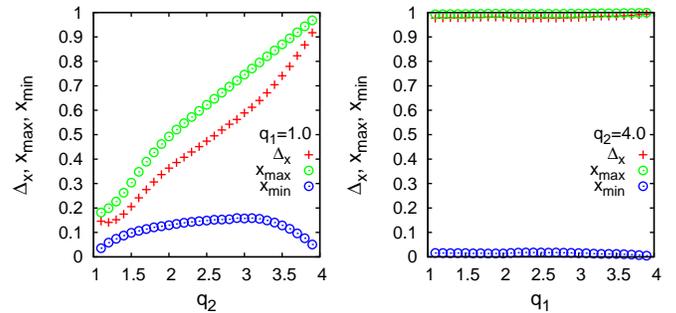}}
\caption{(Color online) TM results: The variations of $\Delta_x$,  $x_{min}$ and  $x_{max}$ with $q_2$ when $q_1=1.0$ (lower panel).
 The variations of $\Delta_x$,  $x_{min}$ and  $x_{max}$ with $q_1$ when $q_2=4.0$ (lower panel). Both the figures are for uniform distribution.
}
\label{rawdyn_ergo}
\end{figure}

\section{Results for TM method}
\subsubsection{Non-chaotic regime}
We choose a  pair of initial values of $x$ which are slightly different and allow them to evolve as a function of time (eq.~{\ref{eq-delta-t}).
Typical evolutions for $q_1=1$ and $q_2=4$ of two initially close $x$ values are shown in figs.~{\ref{rawdyn_tm}a, {\ref{rawdyn_tm}b.
Although $x$ does not attain a fixed point value, we find that $\Delta_t$ indeed goes to zero in an exponential manner
for chosen values of $q_1$ and $q_2$ (e.g. for $q_1 =1, q_2 = 4$; see figs.~{\ref{dyn_in_tm}a, \ref{dyn_in_tm}b) signifying regular  or nonchaotic  behaviour.
However, here we find that $\Delta_t$ varies in a nonlinear manner with 
$\Delta_0$; precisely, $\Delta_t \propto \sqrt{\Delta_0}\exp(\lambda t)$, whereas for non-random logistic map it is $\Delta_t \propto {\Delta_0}\exp(\lambda t)$.
The Lyapunov exponent $\lambda$ depends strongly  on $q_1,q_2$; it shows an increase as
$q_2 - q_1$ is decreased. The value of the Lyapunov exponent is $\lambda \backsimeq -0.186$ for uniform distribution for $1 \leq a_t \leq 4.$

 $\Delta_t$  has  been studied for asymmetric triangular distribution for $1\leq a_t \leq 4$ (Fig.~{\ref{dyn_in_tm}c).
Here the Lyapunov exponent remains constant upto $a_p \approx 1.7$ and decreases with $a_p$ for higher values (Fig.~{\ref{dyn_in_tm}d).
This signifies that as $a_p$ increases, $\Delta_t$ vanishes in a slower manner.

\begin{center}
\begin{table}[h]
\caption{Table for steady state value of $x$.}
\begin{tabular}{|l|l|l|}
\hline
Distribution& Theor. value& Actual value\\
& $\langle x \rangle=\frac {\bar a -1}{\bar a}$&$\langle x_{t \to \infty}\rangle$\\
 \hline
Uniform, $1\leq a\leq 4$&$0.600$ & $0.499$\\ 
Uniform, $3\leq a\leq 4$&$0.714$ & $0.636$\\ 
Symmetric triangular, $1\leq a\leq 4$&$0.600$ & $0.549$\\ 
Symmetric triangular, $3\leq a\leq 4$&$0.714$ & $0.650$\\

\hline
\end{tabular}
\label{TMT}
\end{table}
\end{center}

\begin{figure}[!h]
\resizebox{90mm}{!}{\includegraphics {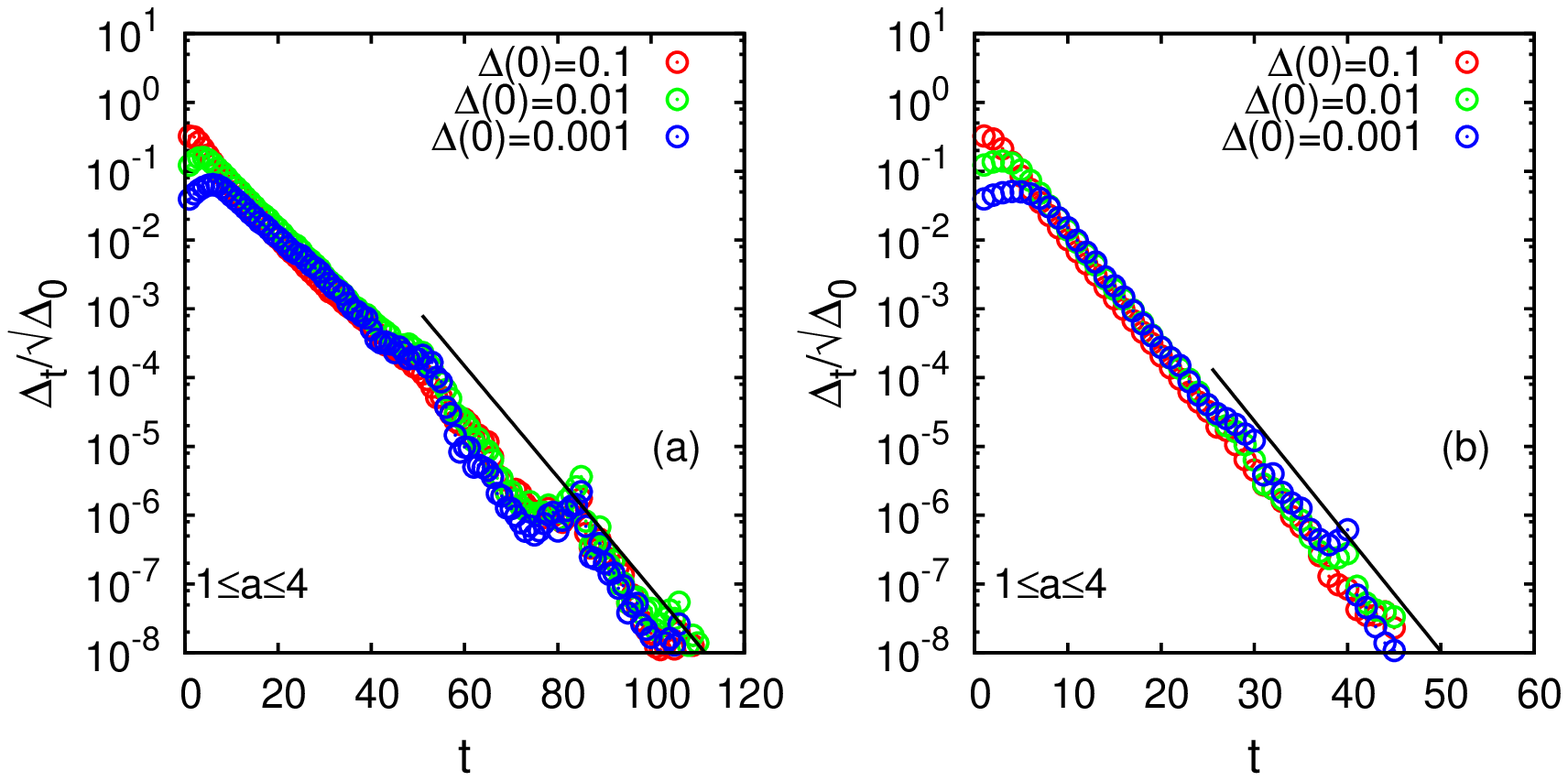}}
\resizebox{90mm}{!}{\includegraphics {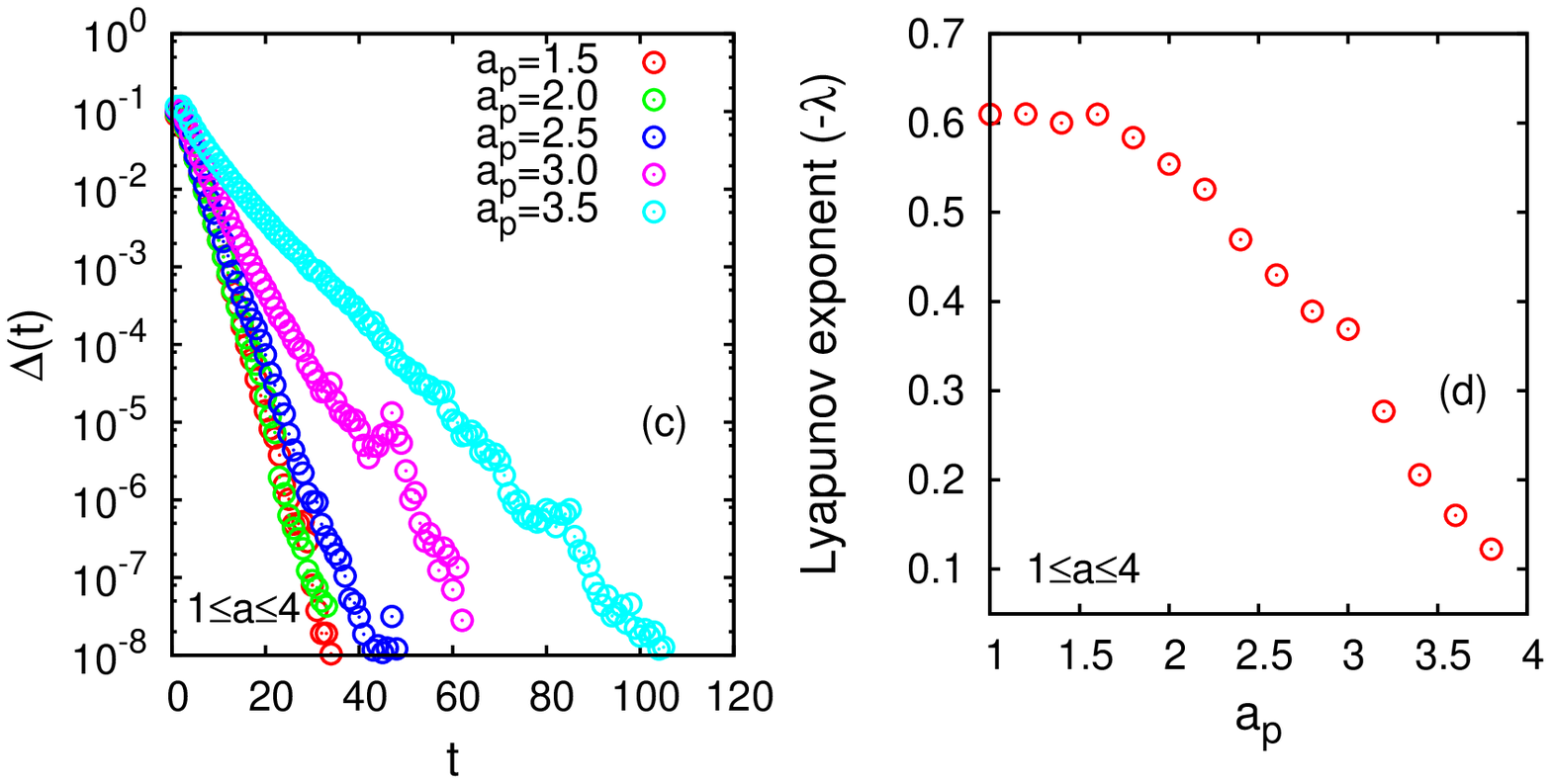}}

\caption{(Color online) TM results: (a) Scaled value of damage $\Delta_t$ against time for uniform distribution,
 (b) scaled value of damage $\Delta_t$ against time for symmetric triangular distribution,
(c) $\Delta_t$ against time for the symmetric triangular distribution on $[1,4]$ with different values of $a_p$ and 
(d) the variation of Lyapunov exponent with the value of  $a_p$ for the  asymmetric triangular distribution on $[1,4]$ peaked at $a_p$.
 \label{dyn_in_tm}
}
\end{figure}
\subsubsection{Chaotic regime}
When $q_2$ is a variable and $q_1=1$, ${\Delta_t \to 0}$ as ${t \to \infty}$ for any value of $q_2$. 
However increasing  $q_1$, we note that $\Delta_{t \to \infty}$ may reach a nonzero value, e.g, when $q_1 = 3$ and $q_2 = 4$ signifying a chaotic behaviour. 
Typical evolutions for $q_1=3$ and $q_2=4$ of two initially close $x$ values are shown in figs.~{\ref{ndyn_in_tm}a, {\ref{ndyn_in_tm}b.
 In this case, the damage saturates to a nonzero value (Figs.~{\ref{ndyn_in_tm}c, \ref{ndyn_in_tm}d main plot). 
We have used here either  a uniform distribution or a symmetric triangular distribution for $a_t$. 
Saturation value of the damage $\Delta_{sat}=\Delta_{t \to \infty}$ has been studied for both the distributions
for different values of $q_1$ and  $q_2$.
One can keep   $q_2$  variable and fix $q_1$ and observe the onset of chaos at a 
threshold value of $q_2$ (Fig.~{\ref{sat_ini_tm} upper panel). Calling this threshold value $q^c_2$, we note that $q^c_2$ is a function of $q_1$ 
and decreases with  $q_1$ which
is expected.  This is  true for both distributions (Fig.~{\ref{sat_ini_tm} upper panel inset). We note that the 
minimum value of $q_1$ for the onset of chaos is $\sim 2.6$ for the uniform distribution and $\sim 2.7$
 for the symmetric triangular distribution. Even for $q_2 <a_c$, where $a_c$ is the threshold value for onset of chaos in the
non random map, one can observe a chaotic region for both the distributions.
\begin{figure}[!h]
\resizebox{90mm}{!}{\includegraphics {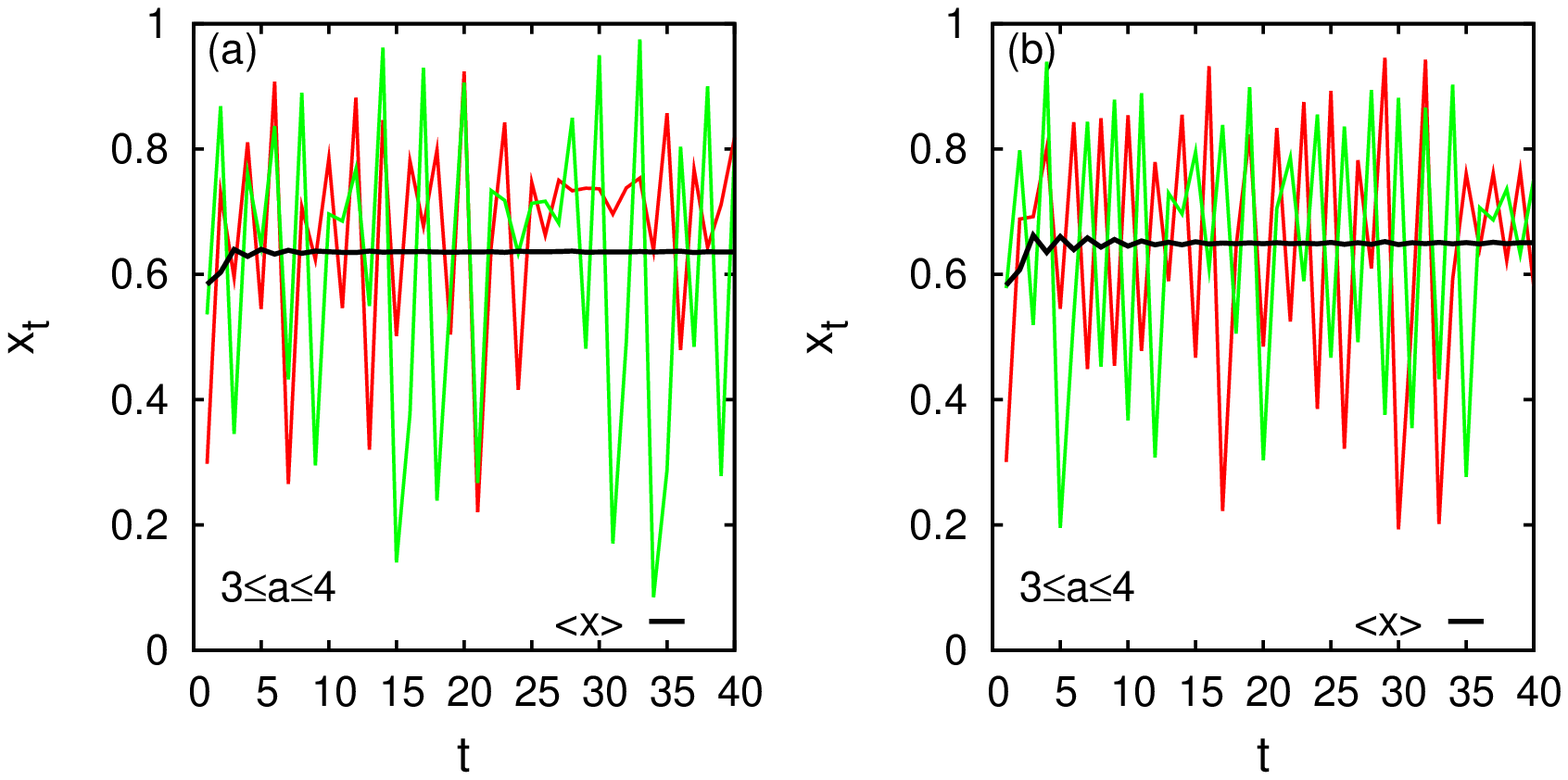}}
\resizebox{90mm}{!}{\includegraphics {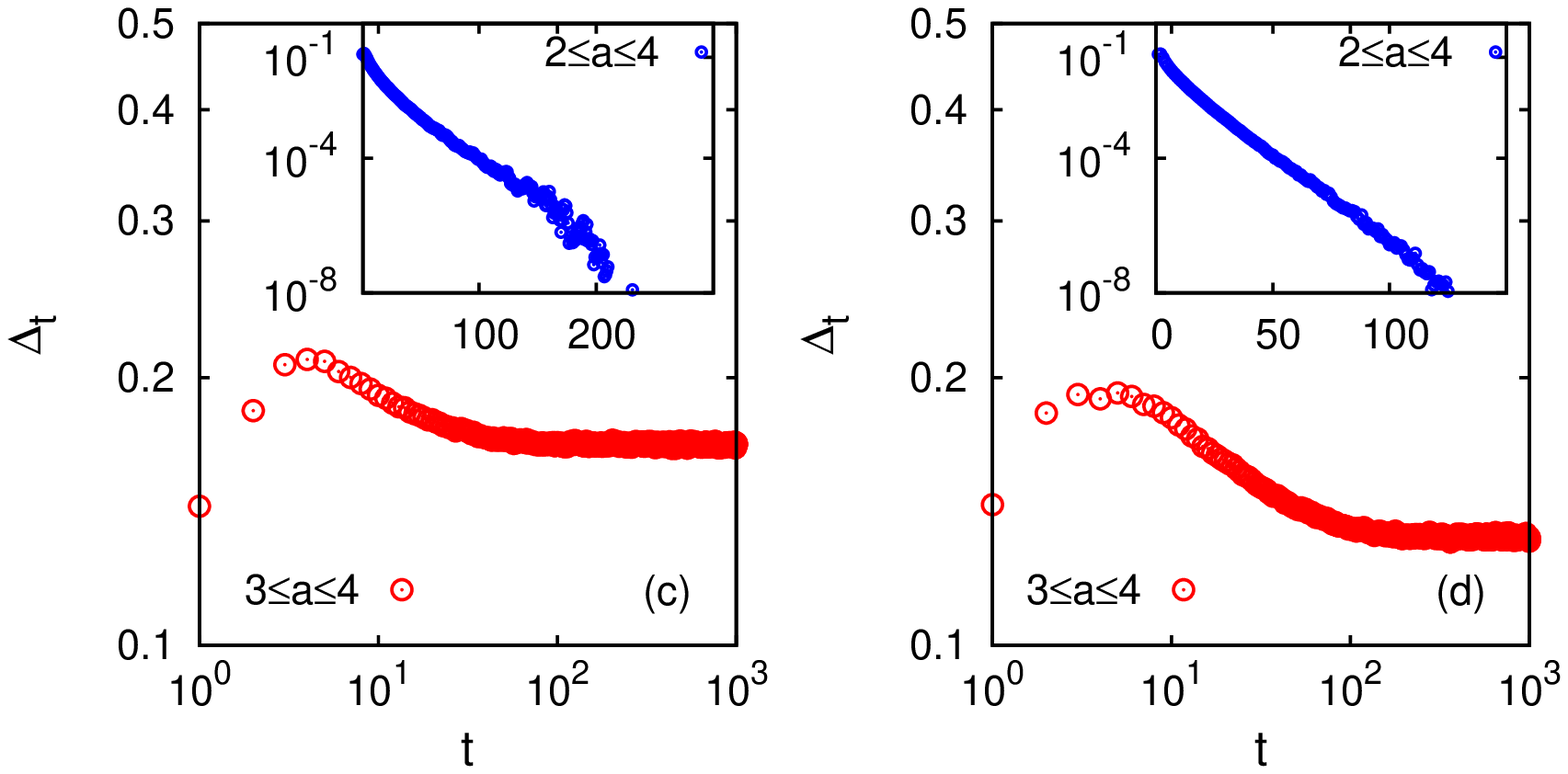}}
 
 \caption{(Color online) TM results: (a) and (b) Two different evolutions of $x(t)$  in the traditional method in the chaotic regime.
The steady state  values  averaged over many configurations also shown.
(c) and (d) $\Delta_t$ against time for $q_2=4$.
Main plot for $q_1=3$ shows saturation of $\Delta_t$ at a nonzero value. Inset for $q_1=2$ shows data when $\Delta_{sat}\to 0$.
  Left panel: uniform distribution; Right panel: symmetric triangular distribution.
 \label{ndyn_in_tm}
}
\end{figure}
Similarly one can keep   $q_1$  variable and fix $q_2$ and observe the onset of chaos at a 
threshold value of $q_1$ (Fig.~{\ref{sat_ini_tm} lower panel). Calling this threshold value $q^c_1$, 
we again note that it is a function of $q_2$ (Fig.~{\ref{sat_ini_tm} lower panel inset).
 What is striking is the presence of a peak  at around $3.6$ which is 
very close to $a_c$. It is found that for $q_2\sim 3.1$, the minimum 
value of $q_1$ required for chaos is $\sim 3.0$. Note that this is the value above
which bifurcations start occurring in the nonrandom map. In fact for $3.1<q_2<3.6$, the threshold value
 $q^c_1$ is weakly dependent on $q_2$ and remains $\sim 3.0$ in the entire region. Above $q_2=3.6$, however, smaller values of $q_1$ allow chaos.


\begin{figure}[!h]
\resizebox{80mm}{!}{\includegraphics {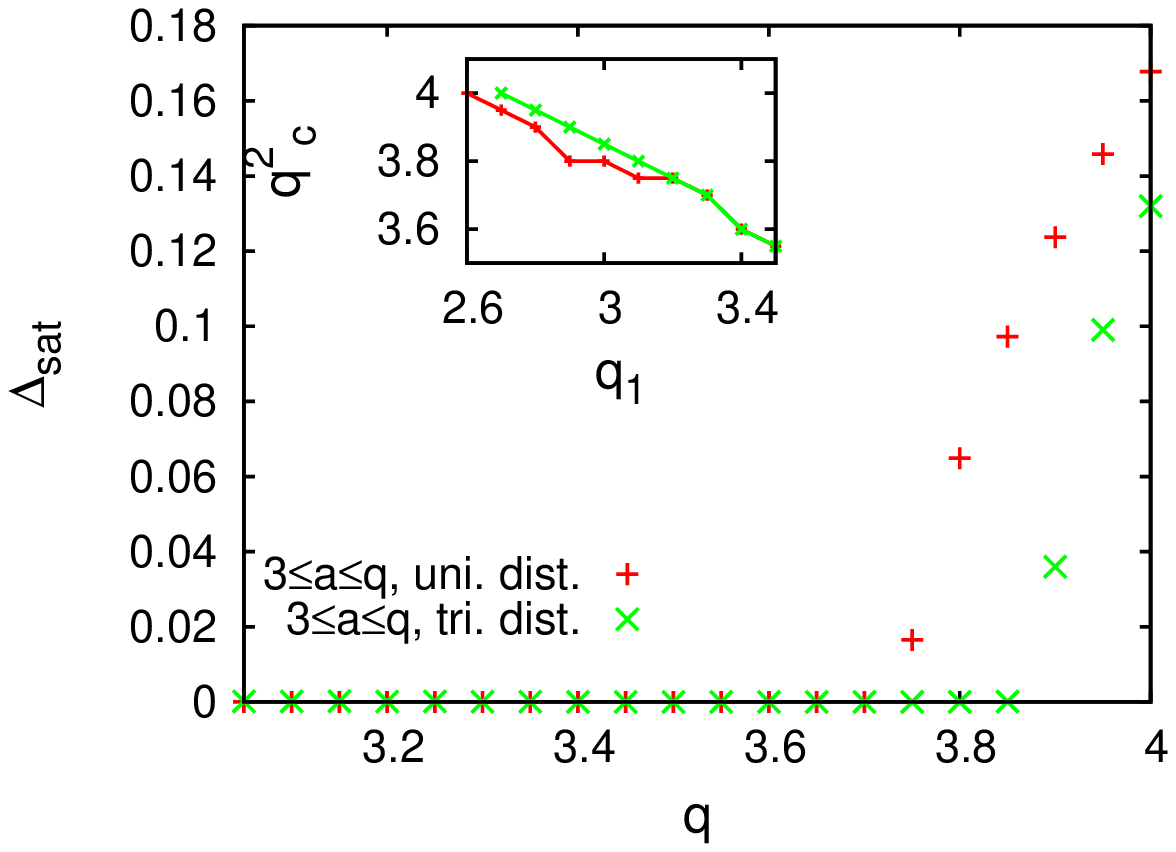}}
\resizebox{80mm}{!}{\includegraphics {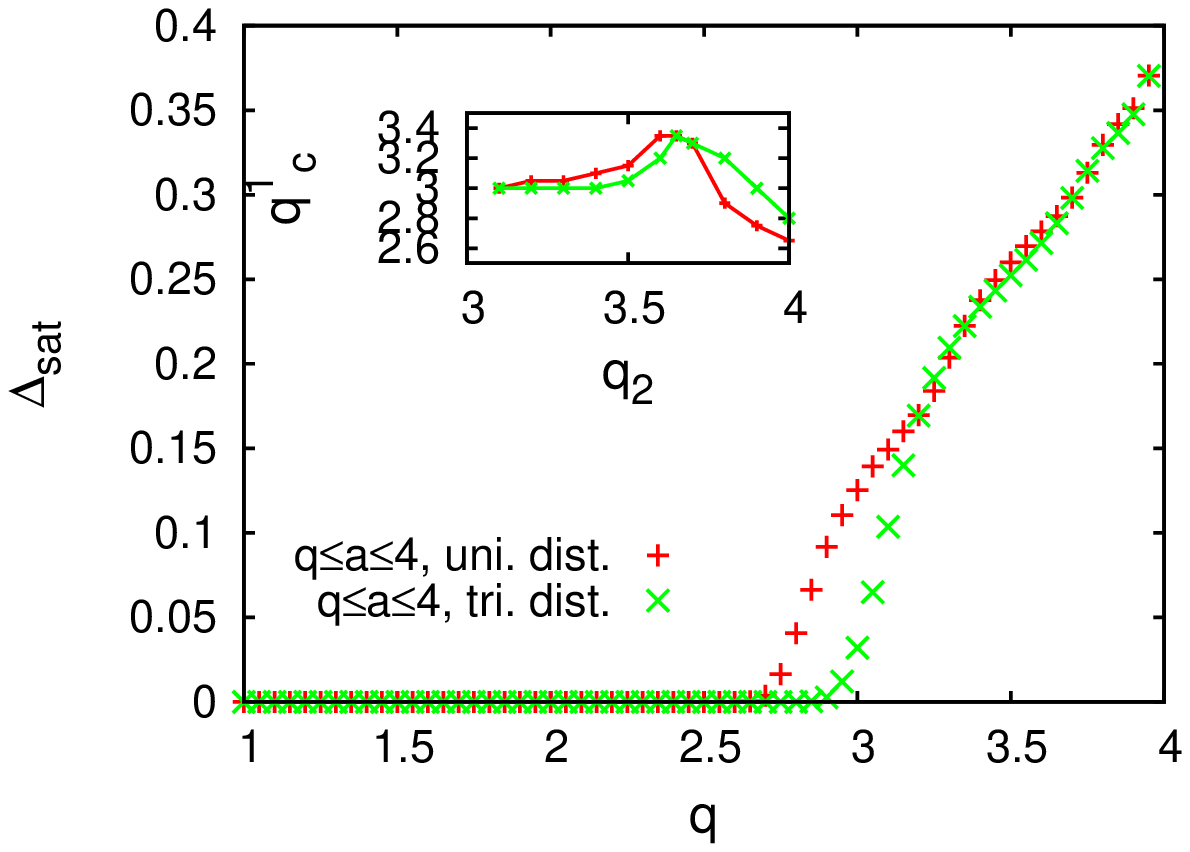}}
\caption{(Color online) TM results: Saturation value of damage $\Delta_{sat}$ shows onset of chaos as $q_1$ is
 fixed and $q_2$ varied (upper panel); $q_2$ is fixed and $q_1$ varied (lower panel). Insets show variation of threshold values $q_2^c$ against $q_1$ (upper panel inset)
and $q_1^c$ against $q_2$ (lower panel inset).
 \label{sat_ini_tm}
}
\end{figure}

\section{Results for the NVN Method}
We next discuss the results for the  NVN method.
In this case, two copies of $x_t$, initially identical, evolve independently, i.e., using different random numbers drawn from
identical distributions. 
Here also the control parameter is chosen from uniform and symmetric triangular distributions.
We find that independent of the values of $q_1$ and $q_2$, the time evolved values of the two copies never converge
as long as $q_1 \neq q_2$.  
Both copies evolve with different values at all times (Fig.~{\ref{rawdyn_nvn}); however, the configuration average of course is the same as that in the TM case.

The damage $\Delta_t$ as a function of time initially  increases  and then  takes a steady value. 
The saturation  value of $\Delta_t$ is nonzero for all the different regions of control parameter
(Fig.~{\ref{dyn_nvn}) as is expected from the evolution of $x_t$. This is again true for both the distributions.
\begin{figure}[!h]
\resizebox{90mm}{!}{\includegraphics {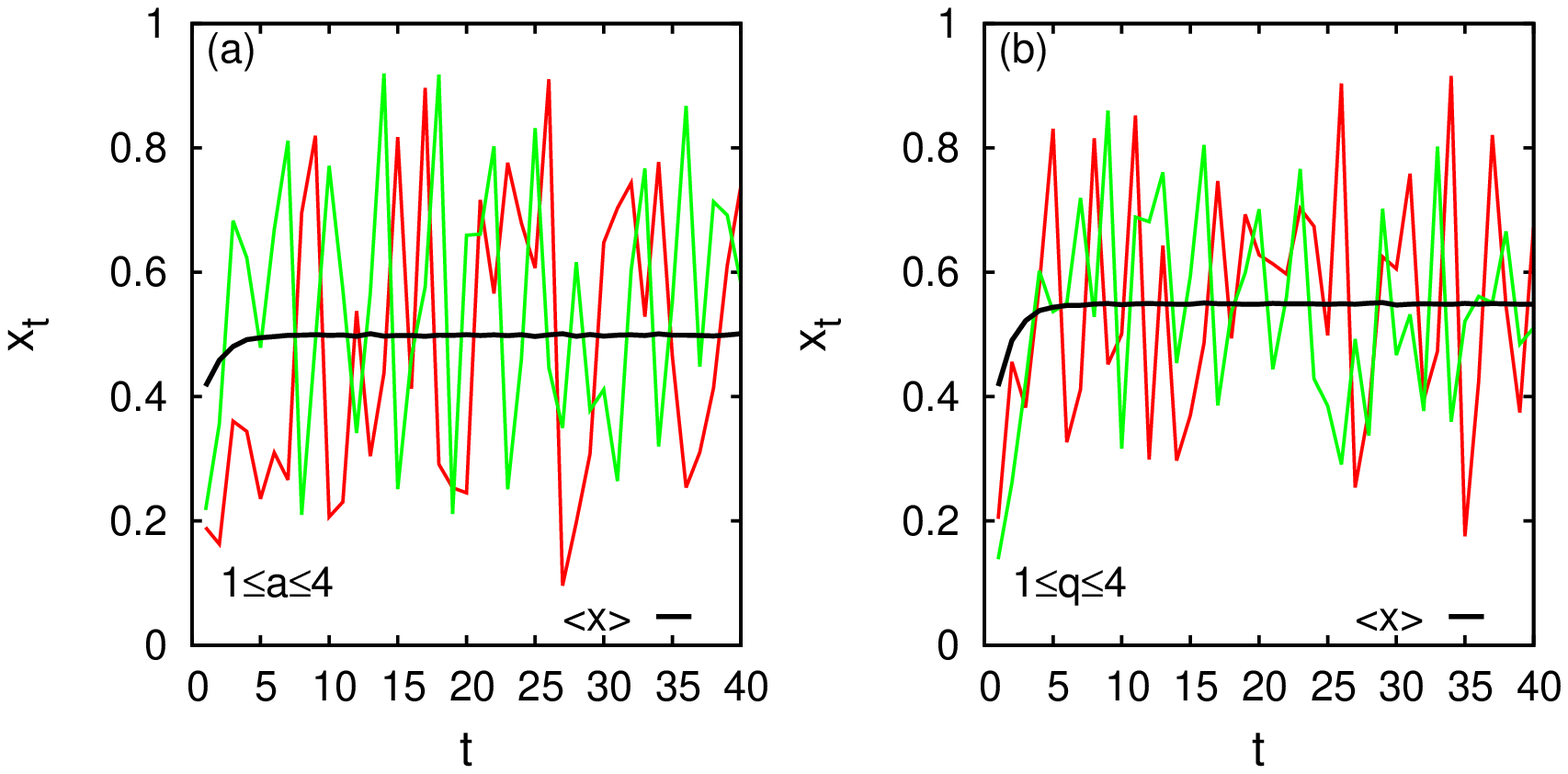}}
\resizebox{90mm}{!}{\includegraphics {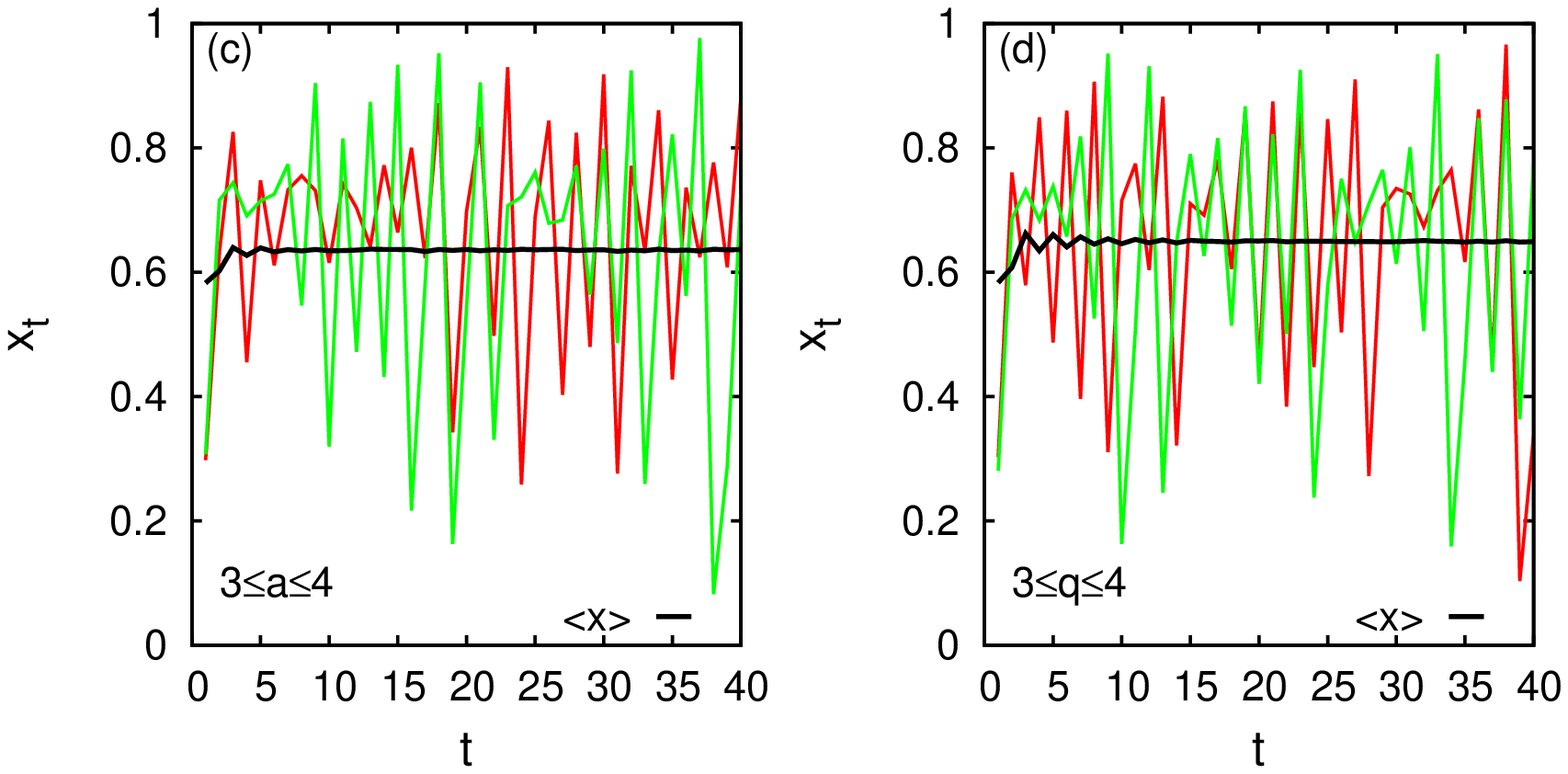}}
\caption{(Color online) NVN results: Two different evolutions of $x(t)$  in the  method. Left panels: uniform distribution; Right panels: symmetric triangular distribution.
The steady state  values  averaged over many configurations also shown.
 \label{rawdyn_nvn}
}
\end{figure}

The saturation value of the damage $\Delta_{sat}$ has been studied. When $q_1$ is fixed and the upper limit $q_2 = q$
is varied   ($ q_1 < q \leq 4$), the value of  $\Delta_{sat}$ increases with $q$ right from $q > q_1$ as there is no threshold value of 
the chaos (Figs.~{\ref{sat_nvn}a, \ref{sat_nvn}b). 

If  we keep  $q_1 = q$  as a variable  ($q \geq 1$) and $q_2$ fixed,  the saturation value $\Delta_{sat}$ shows an interesting behaviour. 
Up to $q_2 \sim 3.0$, it  decreases with $q$. At a critical value of $q_2 \sim 3.5$, it shows a non-monotonic behaviour, with a
sharp rise close to $q=3.0$ before decreasing to zero at $q=q_2$. Beyond this critical value of $q_2$, the initial decrease in $\Delta_{sat}$ 
becomes less prominent and it increases right up to $q=q_2$ indicating there is a sharp discontinuity of $\Delta_{sat}$ at $q = q_2$ (Figs.~{\ref{sat_nvn}c, \ref{sat_nvn}d). 
This behaviour is like that in TM (Fig.~{\ref{sat_ini_tm} main plot), however the difference is, one has an onset of chaos in TM such that 
$\Delta_{sat}$ increases from zero while in NVN, it increases from a nearly constant non-zero value. 

\begin{figure}[!h]
\resizebox{90mm}{!}{\includegraphics {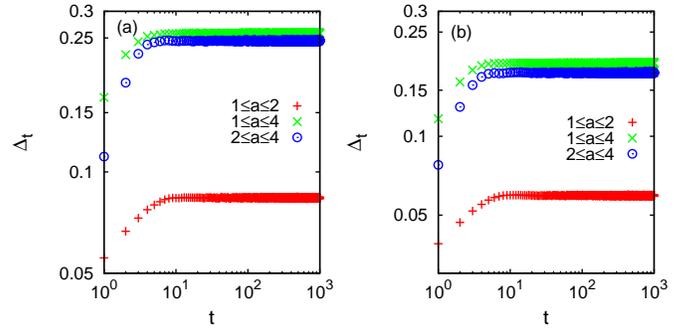}}
\caption{(Color online) NVN results:
 Damage as a function of time for
different regions of control parameter.  Left panel: uniform distribution; Right panel: symmetric triangular distribution.
  \label{dyn_nvn}
}
\end{figure}

\begin{figure}[!h]
\resizebox{90mm}{!}{\includegraphics {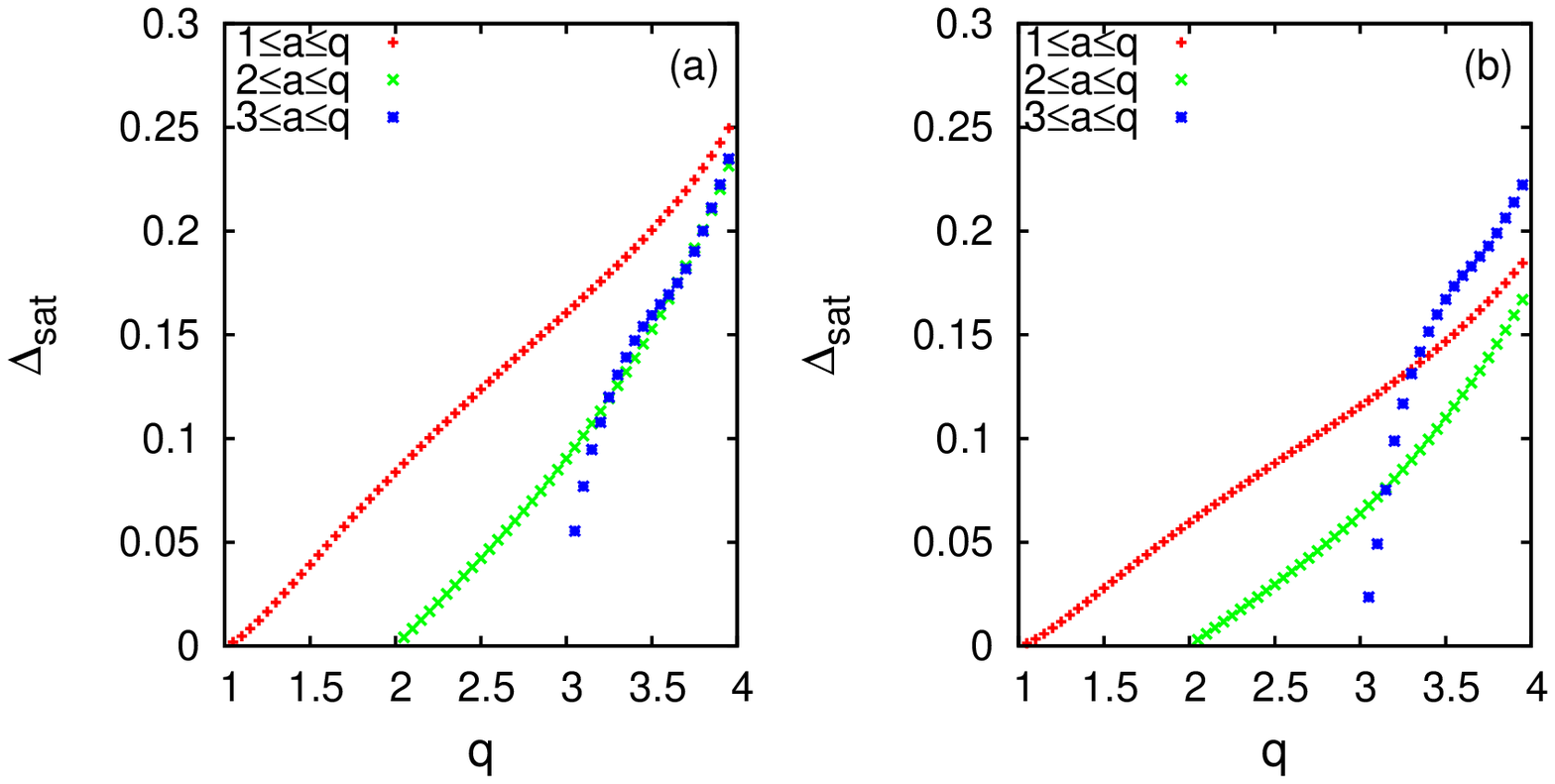}}
\resizebox{90mm}{!}{\includegraphics {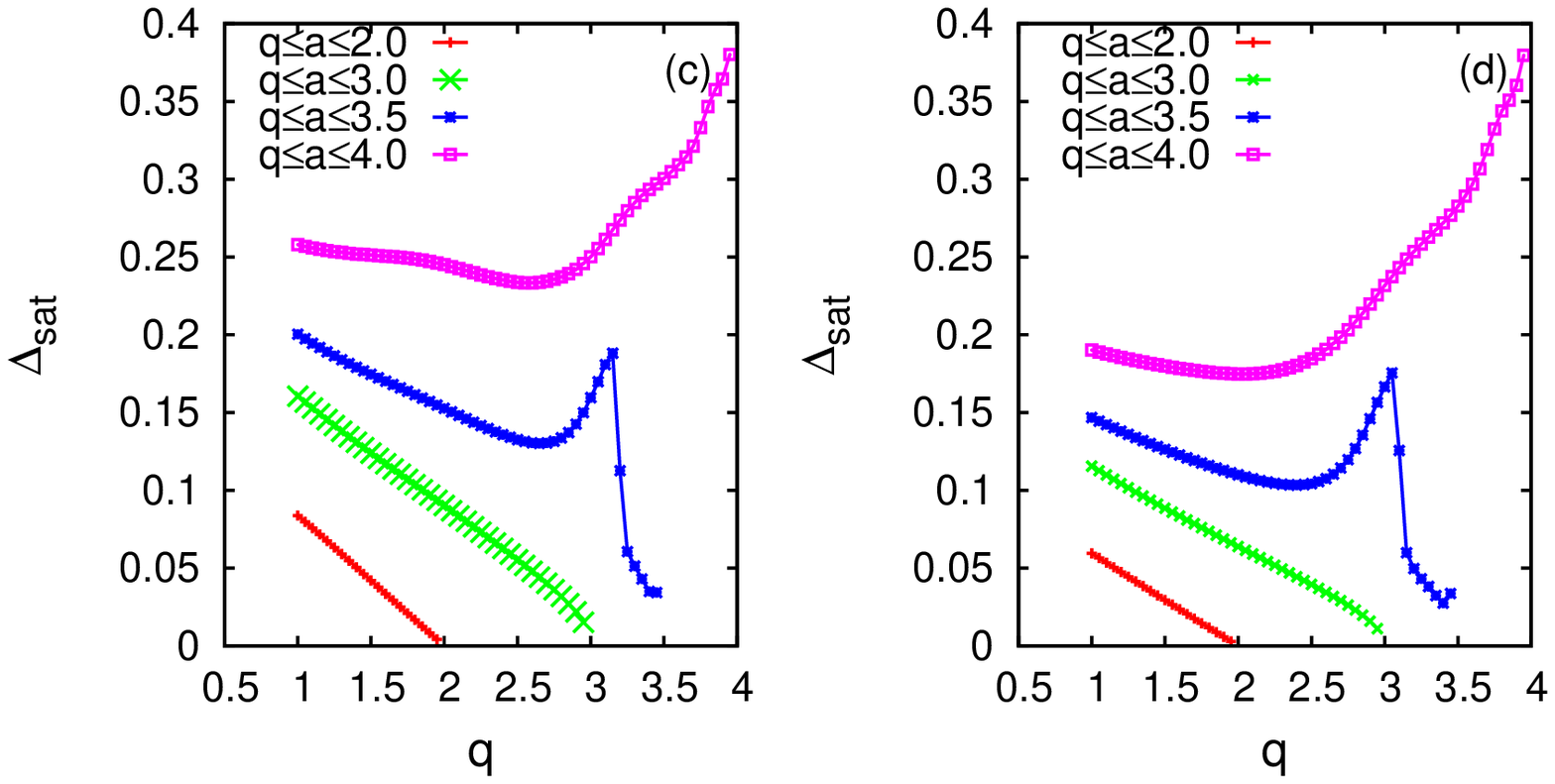}}
\caption{(Color online) NVN results: (a) and (b) Saturation value of damage $\Delta_{sat}$ when $q_1$ is
 fixed and $q_2$ varied (upper panel). (c) and (d) Saturation value of damage $\Delta_{sat}$ when $q_2$ is fixed and $q_1$ varied (lower panel). Left panels: uniform distribution; Right panels:  symmetric triangular distribution.
 \label{sat_nvn}
}
\end{figure}

One can estimate the maximum possible value of $\Delta_{sat}$  assuming two completely 
uncorrelated maps as: 

\begin{eqnarray*}
\Delta_{est}^2 =\int_{0}^1 \! \int_{0}^1 (x_{1}-x_{2})^2 P(x_1) P(x_2)\,dx_{1}\,dx_{2}\\ 
=\int_{0}^1 \! \int_{0}^1 (x_{1}^2-x_{2}^2-2x_1 x_2 )\,dx_{1}\,dx_{2}\ \
\end{eqnarray*}
where $P(x_1)$, $P(x_2)$ denote distribution of $x_1$ and $x_2$. Assuming $P(x_1)$ and $P(x_2)$ to be uniform,  $\Delta^2_{est}=1/6$.
Therefore the expected value of $\Delta_{est}=\sqrt{1/6} \approx 0.41$ for uncorrelated maps.
$\Delta_{sat}$ is indeed less than $\Delta_{est}$ for both NVN and TM.
\section{Summary and discussion}
In summary, we have studied the behaviour of random logistic maps where the parameter $a$ in eq. \ref{eq-logistic} is a random variable. $x_t$
shows semi or fully ergodic behaviour for such maps, however $\langle x_{t \to \infty}\rangle$ attains saturation values which differ from the theoretical mean field values as given by eq. \ref{non_random}.
The deviations are less in case of a symmetric triangular distribution as it has less variance (Table I).

It is known that randomness in linear systems may give rise to chaos \cite{yu}. In the present model, 
we can identify nontrivial nonchaotic behaviour even with both randomness and nonlinearity.
In the nonchaotic
regime we find the unconventional behaviour $\Delta_t \sim \sqrt{\Delta_0} \exp(\lambda t$). 
Here, one can estimate the Lyapunov exponent $\lambda$. We observe that $\lambda$ shows nonuniversality in the sense it
shows strong dependence on the asymmetry of the distribution which may be quantified by $|a_p-(q_1+q_2)/2|$.

Onset of chaos is noted in the traditional
method at threshold values of $q_1$ ($q_2$) which are dependent on $q_2$ ($q_1$). Minimum values for onset of chaos is found to
be $q_1\sim 2.6$ (for the uniform distribution) and $q_1\sim 2.7$ (for symmetric triangular distribution) when $q_2=4.0$. On the other hand, for $q_2$, 
the corresponding minimum value is $\sim 3.1$ with $q_1\sim 3.0$ for both the distributions.
The most striking result is 
even when  $q_2<a_c$, we obtain chaotic region for  $q_1> 3.0$ for both the distributions (Fig.~{\ref{sat_ini_tm}).


In NVN, no threshold values of $q_1$, $q_2$ are obtained. Chaos occurs for all $q_2>q_1$. However $\Delta_{sat}$ shows 
interesting variation with $q_1$ and $q_2$ (Fig.~{\ref{sat_nvn}). In general, $\Delta_{sat}$ increases as $q_2$ increases but
$\Delta_{sat}$ is not trivially dependent on $q_2-q_1$ and hence  we conclude that $\Delta_{sat}$ is a nontrivial function of both $q_1$ and $q_2$.

Although the maps are random, certain effects of the nonrandom maps seem to be present in the TM results, e.g., the peak of $q_1^c$ occurs at $q_2\approx a_c$
and minimum value of $q_1^c$ for $q_2<a_c$ is  $\sim 3.0$ where bifurcation starts occurring in the non random case. Another point that needs to be mentioned is that 
it has not been possible to estimate Lyapunov exponents in the chaotic regime as saturation values are attained within very short times.

As had been observed earlier \cite{machta,khaleque}, the TM and NVN methods yield completely different results. As in the case of damage 
spreading in opinion dynamics model \cite{khaleque}, here too we find that the chaotic regime is obtained for any nonzero value of $q_2-q_1$ in the NVN method.

Acknowledgements: 
AK acknowledges financial support from UGC sanction no. UGC/960/JRF(RFSMS). PS acknowledges financial support from CSIR 
project.
{}
\end{document}